\documentclass[12pt,onecolumn,draftcls]{IEEEtran}
\usepackage{cite}
\usepackage{url}
\usepackage{graphicx}
\usepackage{epsfig}
\DeclareGraphicsExtensions{.pdf,.jpg,.png,.eps}
\usepackage{subfigure}
\usepackage{url}
\newcommand{\figref}[1]{\figurename~\ref{#1}}
\usepackage{verbatim,amsmath,amssymb}
\usepackage{array}
\begin{document}
\title{Latent Sentiment Detection in Online Social Networks: A Communications-oriented View}
\author{\IEEEauthorblockN{Vinay Uday Prabhu$^{1,2}$ , Rohit Negi$^{1}$ and Miguel Rodrigues$^{3}$ \\}
\IEEEauthorblockA{$^{1}$Electrical and Computer Engineering Department, Carnegie Mellon University \\
$^{2}$Departamento de Ciencia de Computadores, University of Porto\\
 $^{3}$Electronic and Electrical Engineering Department, University College London \\
Email:vinayprabhu@cmu.edu, negi@ece.cmu.edu, m.rodrigues@ucl.ac.uk }
 }

%\author{Vinay~Uday~Prabhu,~\IEEEmembership{Student Member,~IEEE,}
%        and~Miguel.R.D.Rodrigues,~\IEEEmembership{Member,~IEEE}
        %and~Jane~Doe,~\IEEEmembership{Life~Fellow,~IEEE}% <-this % stops a space
%\thanks{Copyright (c) 2010 IEEE. Personal use of this material is permitted. However, permission to use this material for any other purposes must be obtained from the IEEE by sending a request to pubs-permissions@ieee.org.

%Vinay Uday Prabhu was with the Instituto de Telecomunica\c{c}\~oes, Departamento de Ciencia de Computadores, Universidade de Porto, Porto - 4169-007, Portugal. He is now with the Department of Electrical and Computer Engineering, Carnegie Mellon University, Pittsburgh, PA 15213-3890, USA.(e-mail: vinaypra@andrew.cmu.edu).

%Miguel.R.D.Rodrigues is with the Instituto de Telecomunica\c{c}\~oes, Departamento de Ciência de Computadores, Universidade de Porto, Porto - 4169-007, Portugal.(e-mail: mrodrigues@dcc.fc.up.pt).
%This work was supported by Funda\c{c}\~ao para a Ciencia e a Tecnologia through the research project PTDC/EEA-TEL/100854/2008 and by  Instituto de Telecomunica\c{c}\~oes through the research project IT/LA/431/2008. 
%A part of the material presented in this paper has been accepted for publication at the IEEE Global Communications Conference - GLOBECOM 2010, Miami, US, December 2010.}% <-this % stops a space

%\thanks{Manuscript received *******; *********.}

\maketitle
\begin{abstract}
In this paper, we consider the problem of latent sentiment detection in Online Social Networks such as Twitter. We demonstrate the benefits of using the underlying social network as an Ising prior to perform network-aided sentiment detection. We show that the use of the underlying network results in substantially lower detection error rates compared to strictly features-based detection. In doing so, we introduce a novel communications-oriented framework for characterizing the probability of error, based on information-theoretic analysis. We study the variation of the calculated error exponent for several stylized network topologies such as the complete network, the star network and the closed-chain network, and show the importance of the network structure in determining detection performance.
\end{abstract}
\section{Introduction}
Online Social Networks (OSN), such as Twitter\footnote{\texttt{ https://twitter.com/} } have come to heavily influence the way people socially galvanize. 
Recent world events such as the Arab Spring, witnessed cascading democratic revolutions characterized by a strong reliance on online social media such as Twitter and Facebook \cite{book_twitter}. Today there are about 554,750,000 registered active Twitter users with about 135,000 new Twitter users signing up everyday. Around 58 million tweets are \textit{tweeted} per day  and the website attracts over	190 million visitors every month \cite{twitter_stats}. Such staggering numbers have turned such OSN into a veritable data gold mine
for organizations and individuals who have a strong social, political or economic interest in maintaining and enhancing their clout and reputation. Therefore, extracting and analyzing the embedded sentiment in the microblogs (or \textit{Tweets}) posted by the tweeters about these organizations or individuals, or specific  issues, products and events related to them or their competitors, is of great interest to them. Of particular interest is the {\em latent sentiment} (as opposed to individual expressed sentiments), which can be either positive or negative with respect to a particular position. We explain this latent sentiment in detail in  Section II.
Strict length restrictions (such as the 140 character-limit per tweet), irregular structure of the microblog content and the usage of sarcasm renders the problem of automatic latent sentiment detection (classifying latent sentiment as positive or negative) from the microblog contents error-prone. As evidenced in literature \cite{ml_sentdetect_1},\cite{ml_sentdetect_2}, sentiment detection has been approached from an engineering perspective with the main focus being on sentiment detection algorithms, followed by empirical performance comparisons using standard datasets such as Stanford Twitter Sentiment (STS) dataset and the Obama-McCain Debate dataset\cite{twitter_data}.We had considered in \cite{slg}, network aided detection of votes in the senate harnessing the joint press release network.
%In this paper, we consider the problem of social-network aided sentiment detection. That is, we use the underlying social network as a graph of sentiment similarity, to improve the performance of latent sentiment detection.  
In this paper, we approach such problems from a relatively scientific perspective.That is, we attempt to answer the following two important questions regarding social-network aided latent sentiment detection. 
\begin{enumerate}
	\item How much benefit can be expected by using the social network for latent sentiment detection? 
	\item How does the social network structure affect the performance of network-aided sentiment detection?
\end{enumerate}
Towards this end, we use a stylized model of latent sentiment detection, based on a Markov Random Field model. We then analyze the performance of the optimal Maximum Aposteriori Probability (MAP) sentiment detector, keeping in mind the underlying social network structure. For this, we are inspired by a communications-oriented viewpoint, where we view the underlying network as providing a \emph{weak channel code}, that transmits one bit of information. Accordingly, we are able to analyze the performance of the sentiment detector, by borrowing tools from information theory. We can then compute and contrast the performance under various stylized social network topologies, thus providing answers to both the questions posed above. Thus, we show that communication theorists can contribute to the growing field of social network analysis.

The rest of the paper is organized as follows.
In Section II, we describe the latent sentiment detection problem, introduce the formal model and motivate its relevance through real-world scenarios based on Twitter. We also specify the optimal MAP sentiment detector. In Section III, we perform a communications-oriented analysis of the MAP sentiment detector to derive an upper bound on the detection error probability, in terms of an error exponent. We also show how the exponent can be evaluated numerically for various stylized topologies such as the complete network, the star network and the chain network. In Section IV, we present numerical results that show how the error exponent depends on the network topology and other social network parameters. We conclude the paper in Section V.
\vspace{-0.5cm}
\section{latent sentiment detection problem}

\subsection{Model}

\begin{figure}[t!]
\centering
\includegraphics[width=75mm,height=60mm]{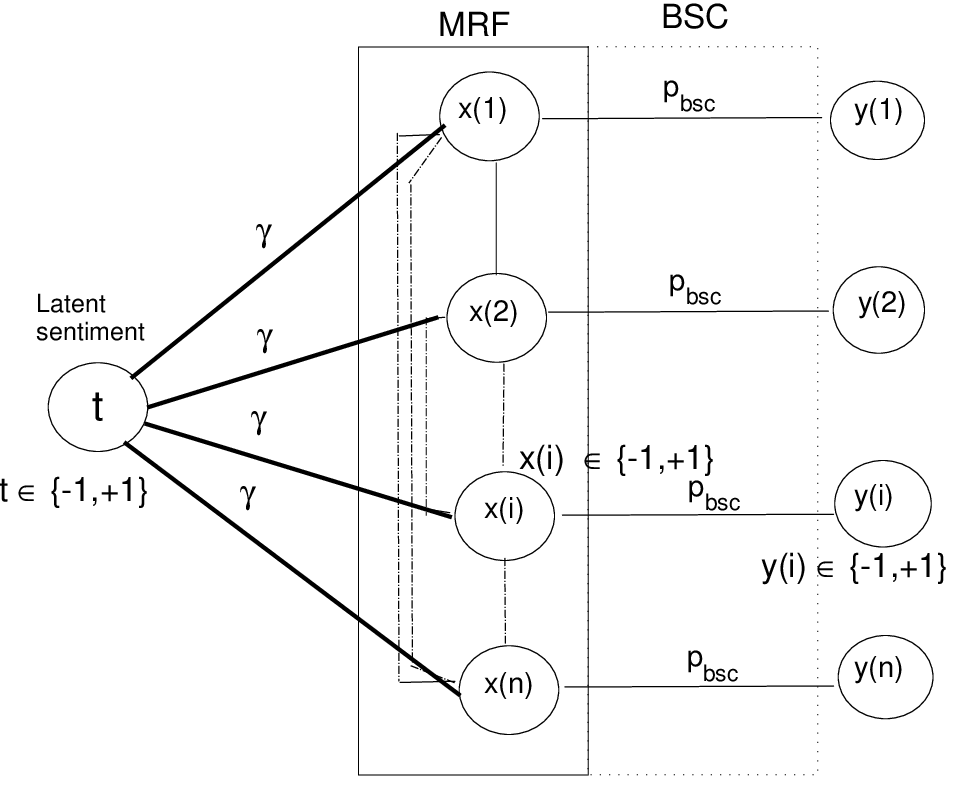}
\caption{Model for latent sentiment detection.}
\label{model}
\end{figure}
The model considered in this paper is shown in \figref{model}. Let ${\bf{x}} \in {\left\{ { - 1, + 1} \right\}^n}$ be the vector of expressed sentiments of the $n $ members of a social network, with $x(i) \in \{-1,+1\}$ being the expressed sentiment of the $i^{th}$ member/node. The social network structure is modeled as an undirected graph $G(V,E)$ characterized by its symmetric adjacency matrix $A$. It may be obtained using the follower/followee relationships, or in some cases, using the $@$-mentions in the tweets\cite{net_sentdetect_2}. The graph is undirected since we will use it to model correlation, rather than influence flows.  The sentiments are assumed to be sampled from an underlying homogeneous Markov Random Field (MRF) \cite{Ising_prior} with unit edge potential and inverse temperature parameter, $\theta$. 
In this paper, we assume $\theta \geq 0$, so that we are restricting ourselves to attractive/ferromagnetic models, which correspond to homophilic networks. In such a ferromagnetic model, the neighboring nodes positively correlate with each other, so that the distribution has  more probability in configurations with similar values on the nodes of the graph. Let $t \in \{-1,+1\}$ indicate the latent sentiment variable which homogeneously influences every node of the network as a local field of strength $\gamma t$. In the absence of any sentiment bias, we assume $t$ to be equi-probably equal to $+1$ or $-1$. 
Thus, the conditional distribution of $\mathbf{x}$ given $t$, can be written as, 
\begin{eqnarray}
p({\bf{x}}|t) & = &  \frac{{\exp \left\{ {\theta {{\bf{x}}^T}A{\bf{x}} + \gamma t{{\bf{e}}^T}{\bf{x}}} \right\}}}{{\sum\limits_{\bf{x}} {\exp \left\{ {\theta {{\bf{x}}^T}A{\bf{x}} + \gamma t{{\bf{e}}^T}{\bf{x}}} \right\}} }}.
\label{px_t}
\end{eqnarray}
Notice that from the communications perspective, $\mathbf{x}$ is a codeword randomly chosen in response to bit $t$.
Let $\mathbf{y}$ be a noisy estimate of $\mathbf{x}$. It may be estimated from the features extracted from the user profiles or could even be the sentiment vector estimated by a given classifier algorithm, such as the ones in \cite{ml_sentdetect_1} and \cite{ml_sentdetect_2}.
While the alphabet of each $y(i)$ can be arbitrary, in this paper, for simplicity, we assume that it is binary $\{-1,+1\}$. 
We model $\mathbf{y}$ to be the output of $n$-identical and independent Binary Symmetric Channels (BSCs) characterized by the equal cross-over probability $p_{bsc}$, with the elements of the true sentiment vector $\mathbf{x}$ as the input. Therefore, the conditional distribution, $p({\bf{y}}|{\bf{x}})$ may be written as,
\begin{equation}
p({\bf{y}}|{\bf{x}}) = \frac{1}{{{c^n}}}  \exp \left\{ {\varepsilon {{\bf{y}}^T}{\bf{x}}} \right\},
\end{equation}
where $\varepsilon= \frac{1}{2} \log \left(\frac{p_{bsc}}{1-p_{bsc}} \right)$ and  $c = 2 \cosh(\varepsilon)$.
The joint distribution of all variables may now be written as,
\begin{eqnarray}
p(t,{\bf{x}},{\bf{y}}) \! \! \! & = & \! \! \!  \frac{1}{{{2 \, Z(t)}}}  \exp \left\{ {\theta {{\bf{x}}^T}A{\bf{x}} + \varepsilon {{\bf{y}}^T}{\bf{x}} + \gamma t{{\bf{e}}^T}{\bf{x}}} \right\}, \qquad
\label{p_txy}  \\
 Z(t) \! \! \!  & =&  \! \! \! {c^n} \sum\limits_{\bf{x}} {\exp \left\{ {\theta {{\bf{x}}^T}A{\bf{x}} + \gamma t{{\bf{e}}^T}{\bf{x}}} \right\}}. 
  \label{Z_t}
 \end{eqnarray}
 \vspace{-0.6cm}
\subsection{Real-world examples from Twitter}
\begin{figure}[t!]
\centering
\includegraphics[width=90mm,height=60mm]{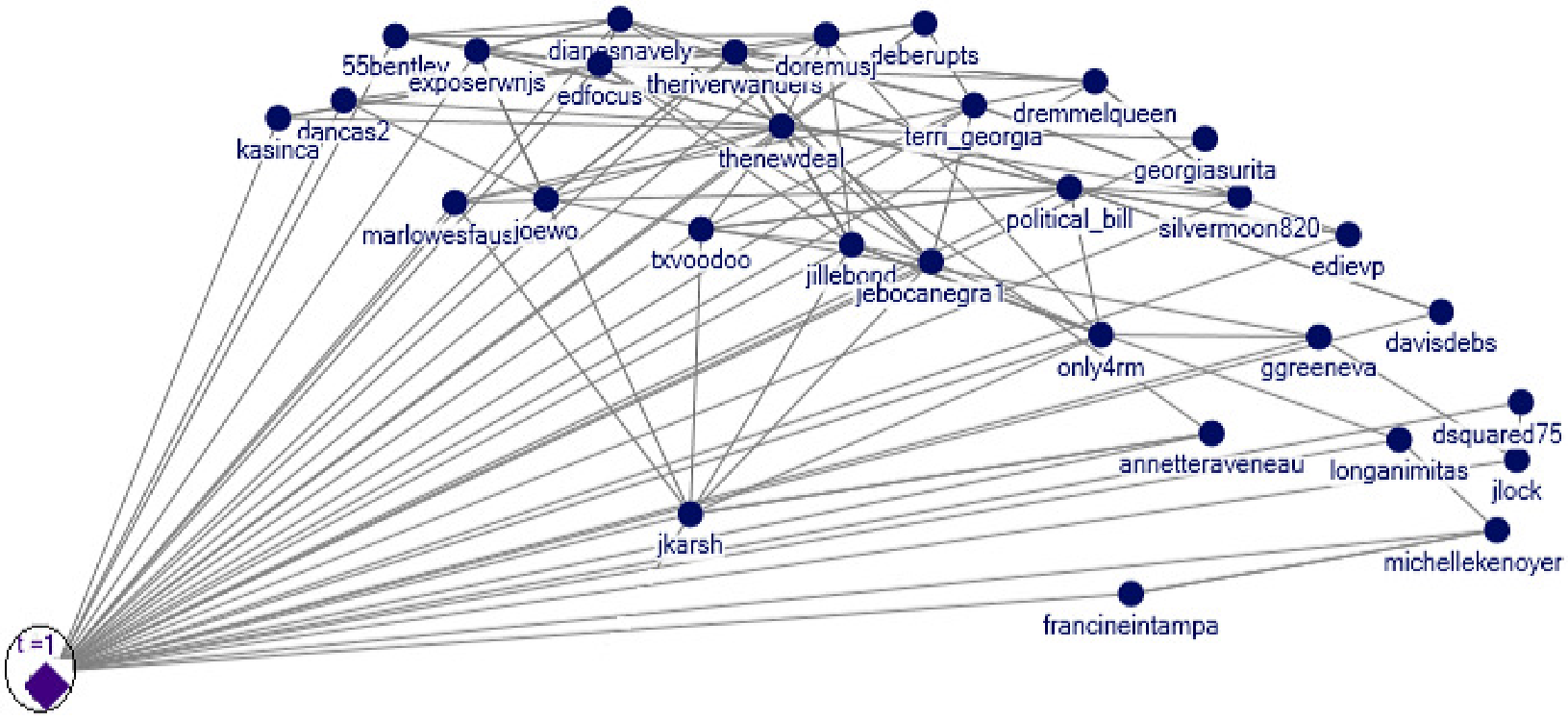}
\caption{The $\#$iloveobamacare network of sentiments.}
\label{obamacare1}
\end{figure}

\begin{figure}[t!]
\centering
\includegraphics[width=90mm,height=60mm]{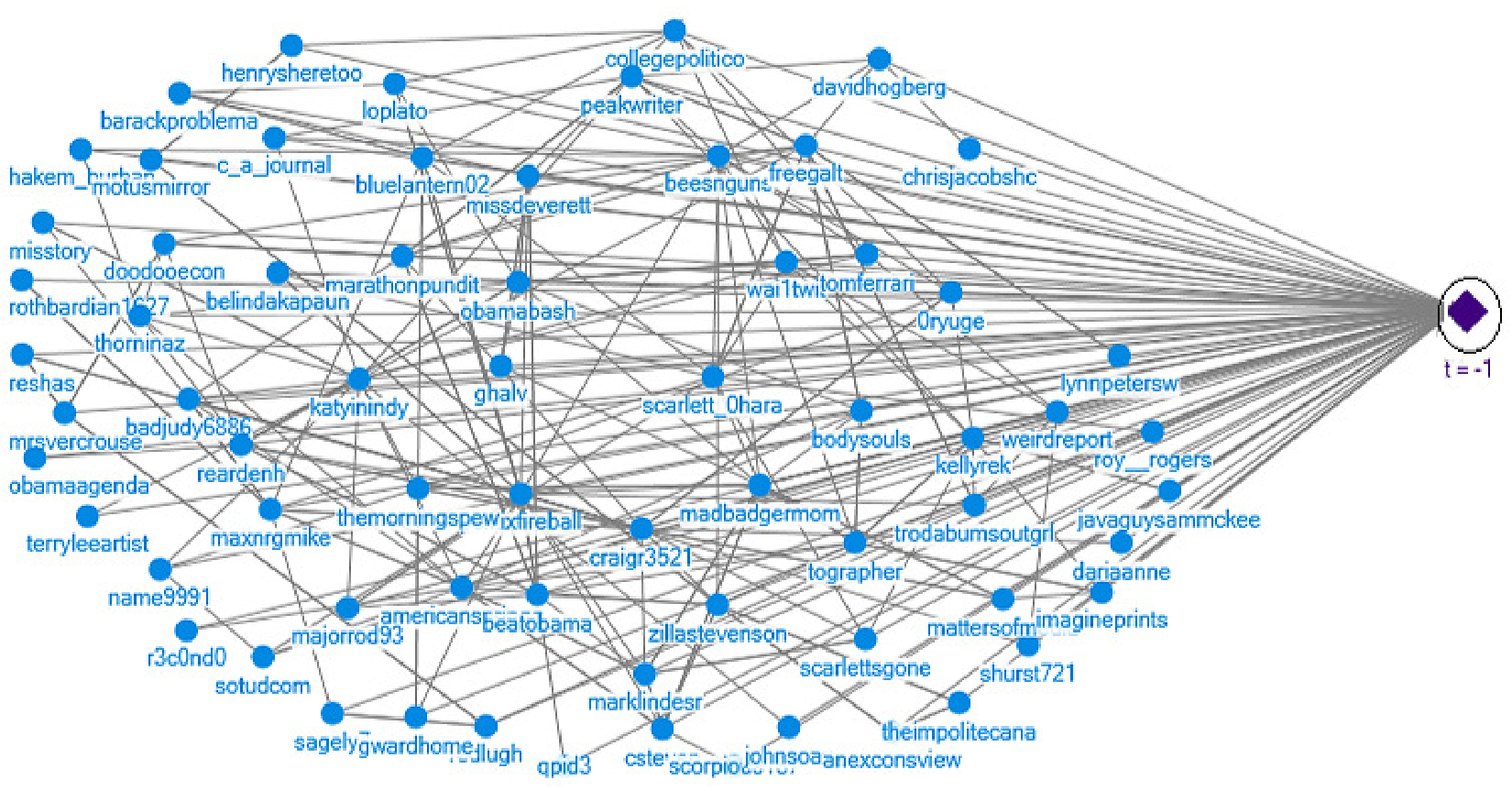}
\caption{The 'anti-supportive' $\#$iloveobamacare network -$t=-1$ scenario}
\label{obamacare2}
\end{figure}

The model \eqref{p_txy} is applicable to  several real-world latent sentiment detection scenarios. We begin by assuming that there exists a latent sentiment ($t\in \{-1,+1\}$), which will cause a certain concrete event in the future. This event may be the passage (or defeat) of a bill in the senate, or an up (or down) movement of the stock market, when $t=+1$ (or $-1$, respectively).  The intention is to predict this real-world event in the present using the expressed sentiments gathered from the \emph{twitterverse} ($\mathbf{y}$ in our model). Thus, it is the same as detecting the value of $t$ (hence the term \textit{latent sentiment detection}).\\ \figref{obamacare1} and \figref{obamacare2} jointly refer to one such scenario where the underlying sentiment $t$ was support (+1) or opposition (-1) to the Patient Protection and Affordable Care Act (PPACA) \cite{obamacare}, nicknamed as \emph{Obamacare}. \figref{obamacare1} represents the social network of liberal-minded follower/followee networks of twitter users, who tweeted in support of the $\#$\texttt{iloveobamacare} hashtag, which was promulgated on twitter to galvanize more support. \figref{obamacare2} represents the follower/followee network of conservative-minded twitter users who attacked the \texttt{$\#$iloveobamacare} hashtag with a series of sharp and sarcastic tweets resulting in what is called \textbf{Hashtag-Hijacking} \cite{obamacare_hijacked}.The national survey conducted by the \texttt{Pew Research Center} and
\texttt{USA TODAY} \cite{pew}, later confirmed the underlying sentiment of support ($t=+1$) to the act amongst liberals and opposition ($t=-1$) to the act amongst the conservatives. \\The goal would be for an automatic sentiment detector to predict $t$ for each network. Of course, in this example, one can do this knowing the political stance of the networks, which is side information. However, automatic detection aims to apply a general method based detector on $\mathbf{y}$ without requiring human intervention through specialized side information.
% As it later revealed in a national survey conducted by the \texttt{Pew Research Center} and \texttt{USA TODAY} \cite{pew}, that $75\%$ of Republican party members opposed the PPACA and believed it would negatively affect the country in the coming years, while $63\%$ of Democrats supported it and thought its impact will be positive.
\vspace{-0.6cm}
\subsection{MAP sentiment detector}
We assume that the network adjacency matrix $A$ and the other system parameters, $\theta, \epsilon, \gamma$ are known. Then, the 
optimal MAP latent sentiment detector $\hat{t}$, which is equal to the Maximum Likelihood (ML) detector in this case, is 
\begin{eqnarray}
\hat t & = &   \left\{
\begin{array}{ll}
 + 1, & l({\bf{y}}) \geq 1 \\
 - 1, &  l({\bf{y}}) < 1
\end{array} \right.
\label{t_hat}
\end{eqnarray}
where, the likelihood $l({\bf{y}})$ is,
\begin{equation}
l({\bf{y}}) = \frac{{\sum\limits_{\bf{x}} {p(t = 1,{\bf{x}},{\bf{y}})} }}{{\sum\limits_{\bf{x}} {p(t =  - 1,{\bf{x}},{\bf{y}})} }}.
\end{equation}
In the next section, we analyze the performance of the optimal MAP sentiment detector \eqref{t_hat}. We perform a communications-inspired analysis of the probability of error of the detector, using which, we seek to understand the contribution of the social network, as well as the role played by the underlying network topology in the performance of the detector.
\vspace{-0.5cm}
\section{Communications-inspired analysis of sentiment detection}
\vspace{-0.1cm}
In this section, we perform an analysis of the error probability of the latent sentiment detector \eqref{t_hat}. By symmetry of the
model, the error probability is,
\begin{eqnarray}
P_e & = & {P_{e|t =  - 1}}  \nonumber \\
& = &  P(l({\bf{y}}) \geq 1 | t=-1) \nonumber \\
& = & \sum\limits_{\bf{y}} {p({\bf{y}}|t =  - 1)} \, {\bf{1}}\left( {l({\bf{y}}) > 1} \right).
\end{eqnarray} 
Here, $\mathbf{1}(A) = 1$ for the event $A$, otherwise it is $0$.$P_e$ is infeasible to calculate for large social network. So, in the next subsection, we present an upper bound for $P_e$. 
\vspace{-0.4cm}
\subsection{$P_e$ upper bound}
The main result of this paper is the following theorem.
\emph{Theorem 1:} For the optimal MAP detector \eqref{t_hat}, an upper bound on the error probability $P_e$ is,
%\begin{eqnarray}
%{P_{e,UB}} & = & \frac{1}{{Z(\theta ,\gamma ){{\left( {\cosh (\varepsilon )} \right)}^n}}} \mathop {\min }\limits_b A(b), \quad \mbox{where},\nonumber \\
%A(b) & = & 
%{\left( {\frac{{\cosh (2b) + \cosh (2\varepsilon )}}{2}} \right)^{n/2}} 
% Z(\theta ,\beta ),
%\label{pe_ub}
%\end{eqnarray}
%where
%$\beta  = \gamma  + \frac{1}{2}\log \left( {\frac{{\cosh (b - \varepsilon )}}{{\cosh (b + \varepsilon )}}} \right)$, $Z(\theta ,\beta ) = \sum\limits_{\bf{x}} {\exp \left\{ {\theta {{\bf{x}}^T}A{\bf{x}} - \beta {{\bf{e}}^T}{\bf{x}}} \right\}}$ and $Z(\theta ,\gamma ) = \sum\limits_{\bf{x}} {\exp \left\{ {\theta {{\bf{x}}^T}A{\bf{x}} - \gamma {{\bf{e}}^T}{\bf{x}}} \right\}}$.\\
\begin{eqnarray}
{P_{e,UB}} & = & \frac{1}{{Z(\theta ,\gamma ){{\left( {\cosh (\varepsilon )} \right)}^n}}} \mathop {\min }\limits_b A(b), \quad \mbox{where,}\nonumber \\
A(b) & = & 
{\left( {\frac{{\cosh (2b) + \cosh (2\varepsilon )}}{2}} \right)^{n/2}} 
 Z(\theta ,\beta ),\nonumber \\
Z(\theta ,\beta )  & = &  \sum\limits_{\bf{x}} {\exp \left\{ {\theta {{\bf{x}}^T}A{\bf{x}} - \beta {{\bf{e}}^T}{\bf{x}}} \right\}}, \nonumber \\
Z(\theta ,\gamma ) & = &  \sum\limits_{\bf{x}} {\exp \left\{ {\theta {{\bf{x}}^T}A{\bf{x}} - \gamma {{\bf{e}}^T}{\bf{x}}} \right\}}.
\label{pe_ub}
\end{eqnarray}

\emph{Proof:} The proof relies on Information-theoretic analysis.(See Appendix).
\vspace{-0.2cm}
\subsection{Computation of the upper bound}
From \eqref{pe_ub}, we see that computation of the upper bound requires computing the partition functions related to the underlying MRF. This is an NP-complete problem in general \cite{z_npc}. However, for certain stylized topologies, such as the complete network, the star network and the (closed) chain network and others, there exist closed form expressions for the partition function.
We compute the partition functions in closed form  for these topologies and list them below.
%\begin{eqnarray}
%{Z_{complete}}(\theta ,\gamma ) & = & \sum\limits_{m = 0}^n {\left( {\begin{array}{*{20}{c}}
%n\\
%m
%\end{array}} \right)} {u_c}^{{{(n - 2m)}^2} - n}{v_c}^{(n - 2m)}  \nonumber \\
%{u_c} &  =  & \exp \left( {\frac{\theta }{2}} \right), \, {v_c} = \exp (\gamma ).   \nonumber\\
%{Z_{chain}}(\theta ,\gamma ) & = &  \lambda _ + ^n + \lambda _ - ^n \nonumber  \\
%\lambda _ \pm ^n   &  =  &   \exp (\theta )\left\{ {\cosh (\gamma ) \pm \sqrt {{{\sinh }^2}(\gamma ) + \exp ( - 4\theta )} } \right\} \nonumber  \\
%{Z_{star}}(\theta ,\gamma )& = &  {u^{n - 1}}\left[ \begin{array}{l}
%{v^n}\sum\limits_{m = 0}^{n - 1} {\left( {\begin{array}{*{20}{c}}
%{n - 1}\\
%m
%\end{array}} \right)} {(uv)^{ - 2m}}  \\
% + {v^{ - n}}\sum\limits_{k = 0}^{n - 1} {\left( {\begin{array}{*{20}{c}}
%{n - 1}\\
%k
%\end{array}} \right)} {\left( {\frac{u}{v}} \right)^{ - 2k}}
%\end{array} \right]  \nonumber \\
%u &= & \exp \left( \theta  \right), \, v = \exp (\gamma ).
%\label{Z_all}
%\end{eqnarray}

\begin{eqnarray}
{Z_{complete}}(\theta ,\gamma ) & = & \sum\limits_{m = 0}^n {\left( {\begin{array}{*{20}{c}}
n\\
m
\end{array}} \right)} {u_c}^{{{(n - 2m)}^2} - n}{v_c}^{(n - 2m)}  \nonumber \\
{u_c} &  =  & \exp \left( {\frac{\theta }{2}} \right), \, {v_c} = \exp (\gamma ).   \nonumber\\
{Z_{chain}}(\theta ,\gamma ) & = &  \lambda _ + ^n + \lambda _ - ^n \nonumber 
\end{eqnarray}
\begin{equation}
\lambda _ \pm ^n  =   \exp (\theta )\left\{ {\cosh (\gamma ) \pm \sqrt {{{\sinh }^2}(\gamma ) + \exp ( - 4\theta )} } \right\}.
\end{equation}
\begin{eqnarray}
{Z_{star}}(\theta ,\gamma )& = &  {u^{n - 1}}\left[ \begin{array}{l}
{v^n}\sum\limits_{m = 0}^{n - 1} {\left( {\begin{array}{*{20}{c}}
{n - 1}\\
m
\end{array}} \right)} {(uv)^{ - 2m}}  \\
 + {v^{ - n}}\sum\limits_{k = 0}^{n - 1} {\left( {\begin{array}{*{20}{c}}
{n - 1}\\
k
\end{array}} \right)} {\left( {\frac{u}{v}} \right)^{ - 2k}}
\end{array} \right]  \nonumber \\
u &= & \exp \left( \theta  \right), \, v = \exp (\gamma ).
\label{Z_all}
\end{eqnarray}
Thus the error probability upper bound \eqref{pe_ub} can be evaluated for the above network topologies, to provide insight into the impact of social network structure on the sentiment detector performance. Note that, via \eqref{pe_ub}, we have reduced the complicated problem of computing an error probability to a problem of calculating an MRF partition function. The partition function calculation is a well researched problem in MRF theory \cite{Z_2},\cite{exact_planar} \cite{Z_3}, and significant effort has been expended in statistical physics and machine learning to compute it for a variety of graphs. Thus, our theorem facilitates importing ideas from that literature to obtain the error probability bound for a variety of graphs.
\vspace{-0.4 cm}
\subsection{Error exponent}
For large networks, an error exponent can be defined as,
\begin{equation}
\alpha  = \mathop {\lim }\limits_{n \to \infty } \inf \left\{ {\frac{{ - \log {P_e}}}{n}} \right\}.
\end{equation}
Using the bound \eqref{pe_ub}, we can show that,
\begin{equation}
\begin{array}{l}
\alpha  \ge \log (\cosh (\varepsilon )) + \frac{{\log (Z(\theta ,\gamma ))}}{n} - \\
\mathop {\min }\limits_b \left[ {\frac{1}{2}\log \left( {\frac{{\cosh (2b) + \cosh (2\varepsilon )}}{2}} \right) + \frac{{\log (Z(\theta ,\beta ))}}{n}} \right].
\end{array}
\label{alpha}
\end{equation}
For the simple case where the network is absent, $\alpha$ can be shown to be exatly,
\begin{equation}
\begin{array}{l}
{\alpha _{iid}} = \log (\cosh (\varepsilon )) + \log (\cosh (\gamma ))\\
\quad \quad \quad  - \frac{1}{2}\log \left( {{{\cosh }^2}(\varepsilon ) + {{\cosh }^2}(\gamma )} \right)
\end{array}.
\label{alpha_iid}
\end{equation}
 Thus, we can use \eqref{alpha_iid} as a benchmark to compare error exponents obtained for various network topologies via evaluation of \eqref{alpha}. In the following section, we perform these comparisons by plotting the variation of the error exponent $\alpha$ derived in \eqref{alpha} and \eqref{alpha_iid} with respect to the model parameters, $\theta$, $\gamma$ and $\varepsilon$.
\vspace{-0.3cm}
\section{Numerical results: Error exponent of different networks}
\begin{figure}[t!]
\centering
\includegraphics[width=90mm,height=60mm]{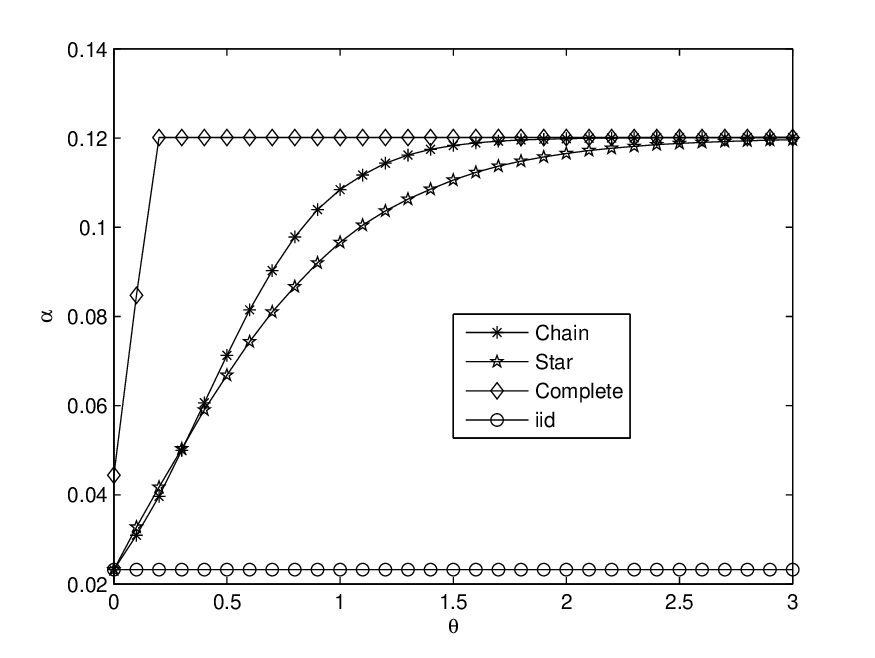}
\caption{Variation of error error exponent ($\alpha$) with $\theta$ for different network topologies}
\label{alpha_theta}
\end{figure}

\begin{figure}[t!]
\centering
\includegraphics[width=90mm,height=60mm]{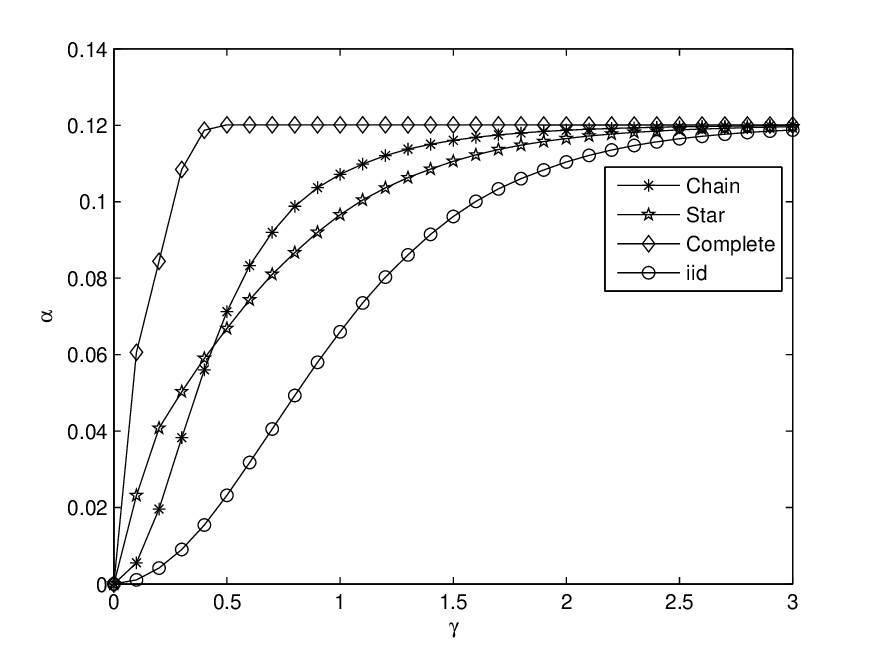}
\caption{Variation of the error exponent ($\alpha$) with $\gamma$ for different network topologies}
\label{alpha_gamma}
\end{figure}

\begin{figure}[t!]
\centering
\includegraphics[width=90mm,height=60mm]{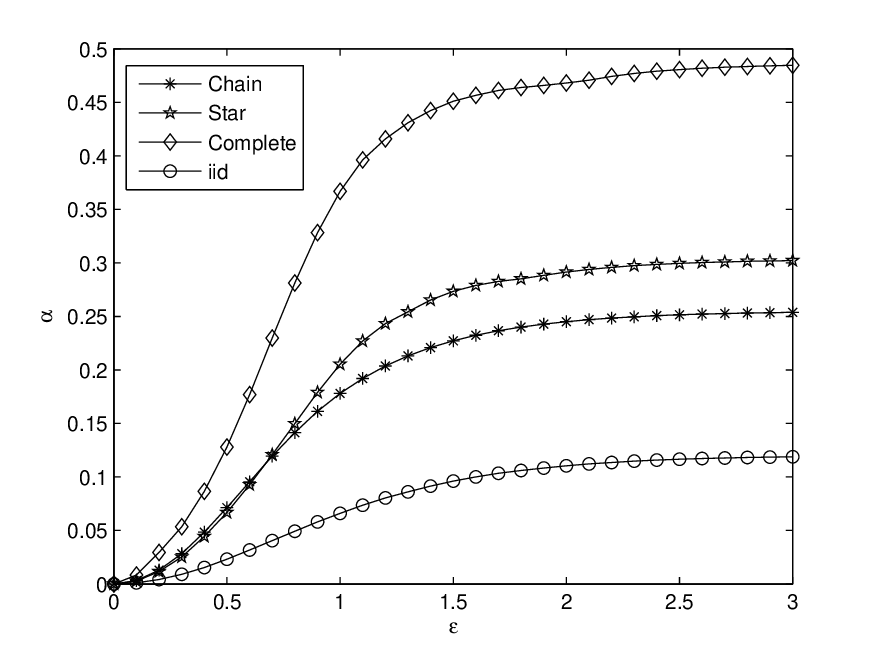}
\caption{Variation of the error exponent ($\alpha$) with $\epsilon$ for different network topologies}
\label{alpha_eps}
\end{figure}

The aim of this section is to answer the two questions raised in the Introduction. First, to show how much improvement can be expected when the network is used for detection and secondly, to understand the effect of network topology on MAP detector performance. We do this by plotting \eqref{alpha}.\\The inverse temperature parameter, $\theta$, captures how strongly  opinions are correlated in the network. Increasing $\theta$ increases the homophily in the OSN, which makes the underlying MRF more strongly ferromagnetic. For example, nodes of a particular political party will be strongly homophilic.This is exploited by the network aided sentiment detector in \eqref{t_hat}, resulting in faster increase of  $\alpha$ for network aided detectors with increasing $\theta$,as shown in \figref{alpha_theta}, with $\varepsilon=0.5$ and $\gamma=0.5$. As seen, $\alpha$ for the no-network (i.i.d.) scenario remains invariant to the change in $\theta$ while for the network-aided  cases, we see $\alpha \to \log(\cosh(\varepsilon))=0.1201$ as $\theta$ increases. The rate of increase of $\alpha$ is clearly the highest for the complete network case when compared to the star and chain networks.Notice also that the complete network reaches the maximum $\alpha$ beyond a certain threshold $\theta$ since it is strongly homophilic.\\The parameter $\gamma$ captures how strongly the latent sentiment influences the opinions of the users/nodes. Practically, large $\gamma$ pertains to a highly emotive situation where the users are strongly influenced to take a particular position even in the absence of strong homophily ($\theta$). Since $\gamma$ is independent of the network effect, both the \texttt{No-net} and the network-aided detectors can harness the effect of increase in $\gamma$ resulting in higher $\alpha$ in both scenarios. However, the combined use of the network as well as $\gamma$ results in higher $\alpha$  for the network aided cases. This behavior is seen in \figref{alpha_gamma}. The other system parameters were held constant ($\varepsilon=0.5$ and $\theta=0.5$). Again, $\alpha \to \log(\cosh(\varepsilon))=0.1201$ as $\gamma$ increases similar to \figref{alpha_theta}.\\Finally, we turn out attention towards \figref{alpha_eps} which shows the variation of $\alpha$ with $\varepsilon$ for $\theta=0.5$ and $\gamma=0.5$. $-\varepsilon$  captures the amount of noise in the BSC in dB (with $\varepsilon \to \infty$ being the zero-noise case). Alternately, $\varepsilon$ is the accuracy of the detector used to obtain estimated individual sentiments, $\mathbf{y}$. As expected, increase in $\varepsilon$ results in higher $\alpha$ for both network-aided and the no-network cases.
In all cases, the error exponent $\alpha$ of the complete network is significantly larger than the star and the chain networks. 

\vspace{-0.5cm}
\section{Conclusion}
In this paper, we have introduced a novel communications-inspired framework for analyzing probability of error of network-aided detection of latent sentiment in Online Social Networks. Through this, we have attempted to provide insight into the role played by the network, specifically the topology, in lowering the probability of error of detection, thereby rigorously characterizing  the worth of the network as an additional information source for sentiment detection. Firstly, we motivate the practical scenarios where the model is applicable and then provide an analysis of the upper bound on the probability of error, or equivalently, the error exponent for large networks. Finally, we plot the variation of this upper bound with respect to model parameters for the complete network, star network and chain network topologies, and show the improvement in performance relative to non-network-aided detection.

\appendix

Following \cite{gallager}, we can upper bound the probability of error using an arbitrary function $0 \le s({\bf{y}}) \le 1$ as, 
\begin{eqnarray}
%{P_{e|t =  - 1}} &\le \sum\limits_{\bf{y}} {p({\bf{y}}|t =  - 1)} {\left( {l({\bf{y}})} \right)^{s({\bf{y}})}};0 \le s({\bf{y}}) \le 1
{P_{e}} &\le & \sum\limits_{\bf{y}} {p({\bf{y}}|t =  - 1)} {\left( {l({\bf{y}})} \right)^{s({\bf{y}})}}
\nonumber \\
&= & \sum\limits_{\bf{y}} {p({\bf{y}}|t =  - 1)}^{\left(1-s({\bf y})\right)} {p({\bf{y}}|t =  + 1)}^{s({\bf y})} \nonumber \\
& \le  & \sum\limits_{\bf{y}} 2  \times {\left( {\sum\limits_{\bf{x}} {p(t =  - 1,{\bf{x}},{\bf{y}})} } \right)^{(1 - s({\bf{y}}))}} \nonumber \\
& & \times \left( {\sum\limits_{\bf{x}} {p(t =  + 1,{\bf{x}},{\bf{y}})} } \right)^{s({\bf y})} \nonumber \\
&  \le & \! \! 2 \sum\limits_{\bf{y}}  {\left( {\sum\limits_{\bf{x}} {\frac{{\exp \left\{ {\theta {{\bf{x}}^T}A{\bf{x}} + \varepsilon {{\bf{y}}^T}{\bf{x}} - \gamma {{\bf{e}}^T}{\bf{x}}} \right\}}}{{Z( - 1)}}} } \right)^{1 - s({\bf{y}})}} \nonumber \\
& & \times {\left( {\sum\limits_{\bf{x}} {\frac{{\exp \left\{ {\theta {{\bf{x}}^T}A{\bf{x}} + \varepsilon {{\bf{y}}^T}{\bf{x}} + \gamma {{\bf{e}}^T}{\bf{x}}} \right\}}}{{Z( + 1)}}} } \right)^{s({\bf{y}})}}. \nonumber
\end{eqnarray}
By optimizing over the function $s({\bf{y}})$, it is possible to write a tighter upper bound,
\begin{eqnarray}
{P_e} & \le & \frac{1}{{{c^n}}} \sum_{\bf y} {\min }_{s ({\bf y})}  \left( {E\left( {\exp \left\{ {+\varepsilon {{\bf{y}}^T}{\bf{x}}} \right\}} \right)} \right)^{(1 - s({\bf{y}}))} \times \nonumber \\
& & \qquad  \left( {E\left( {\exp \left\{ {- \varepsilon {{\bf{y}}^T}{\bf{x}}} \right\}} \right)} \right)^{s({\bf{y}})} \nonumber
\end{eqnarray}
by using \eqref{Z_t}. Here,
 $E \left(\cdot\right)$ denotes expectation with respect to ${\bf x}$, conditioned on $t=-1$ (see \eqref{px_t}).
Using
%\begin{align}
%{P_e} \le \frac{1}{{{c^n}}}\sum\limits_{\bf{y}} {\mathop {\min }\limits_{s({\bf{y}})} & \left\{ \begin{array}{l}
%{\left( {E\left( {\exp \left\{ {\varepsilon {{\bf{y}}^T}{\bf{x}}} \right\}} \right)} \right)^{(1 - s({\bf{y}}))}}\\
% \times {\left( {E\left( {\exp \left\{ { - \varepsilon {{\bf{y}}^T}{\bf{x}}} \right\}} \right)} \right)^{s({\bf{y}})}}
%\end{array} \right\}},
%\end{align}
%where $E()$ represent the expectation taken with respect to the distribution of ${\bf x}$.
%$p({\bf{x}}) = \frac{{\exp \left\{ {\theta {{\bf{x}}^T}A{\bf{x}} - \gamma {{\bf{e}}^T}{\bf{x}}} \right\}}}{{\sum\limits_{\bf{x}} {\exp \left\{ {\theta {{\bf{x}}^T}A{\bf{x}} - \gamma {{\bf{e}}^T}{\bf{x}}} \right\}} }}$.

\begin{align}
\mathop {\min }\limits_{0 \le s \le 1} \left( {a_1^sa_2^{(1 - s)}} \right) \le \frac{{{a_1}{e^s} + {a_2}{e^{ - s}}}}{{{e^s} + {e^{ - s}}}} \le \frac{{{a_1}{e^s} + {a_2}{e^{ - s}}}}{2}, \nonumber
\end{align}
we continue the upper bound as,
\begin{align}
{P_e} \le \frac{1}{{{2c^n}}} \sum_{\bf y} {\min }_{s ({\bf y})} \Bigg\{ & \left( {E\left( {\exp \left\{ {+\varepsilon {{\bf{y}}^T}{\bf{x}}} \right\}} \right)} \right)^{- s({\bf{y}})} + \nonumber \\
& \left( {E\left( {\exp \left\{ {- \varepsilon {{\bf{y}}^T}{\bf{x}}} \right\}} \right)} \right)^{s({\bf{y}})} \Bigg\} \nonumber
\end{align}
%$\mathop {\min }\limits_{0 \le s \le 1} \left( {a_1^sa_2^{(1 - s)}} \right) \le \frac{{{a_1}{e^s} + {a_2}{e^{ - s}}}}{{{e^s} + {e^{ - s}}}} \le \frac{{{a_1}{e^s} + {a_2}{e^{ - s}}}}{2}$ in the above equation results in the following upper bound,
%\begin{equation}
%{P_e} \le \frac{{\sum\limits_{\bf{y}} {\mathop {\min }\limits_{s({\bf{y}})} \left\{ \begin{array}{l}
%\left( {E\left( {\exp \left\{ { - \varepsilon {{\bf{y}}^T}{\bf{x}}} \right\}} \right)} \right){e^{s({\bf{y}})}}\\
% + \left( {E\left( {\exp \left\{ {\varepsilon {{\bf{y}}^T}{\bf{x}}} \right\}} \right)} \right){e^{ - s({\bf{y}})}}
%\end{array} \right\}} }}{{2{c^n}}}
%\end{equation}
and by writing $s({\bf{y}}) =  - {{\bf{b}}^T}{\bf{y}}$, we obtain 
\begin{align}
{P_e} \le \frac{1}{{{2c^n}}} \sum_{\bf y} {\min }_{\bf b} \Bigg\{ &  {E\left( {\exp \left\{ { - {{\bf{y}}^T}\left( {{\bf{b}} + \varepsilon {\bf{x}}} \right)} \right\}} \right)}  + \nonumber \\
&  {E\left( {\exp \left\{ {  {{\bf{y}}^T}\left( {{\bf{b}} + \varepsilon {\bf{x}}} \right)} \right\}} \right)}  \Bigg\} \nonumber 
\end{align}

%\begin{equation}
%{P_e} \le \frac{1}{{2{c^n}}}\sum\limits_{\bf{y}} {\mathop {\min }\limits_{\bf{b}} \left\{ \begin{array}{l}
%E\left( {\exp \left\{ { - {{\bf{y}}^T}\left( {{\bf{b}} + \varepsilon {\bf{x}}} \right)} \right\}} \right)\\
% + E\left( {\exp \left\{ {{{\bf{y}}^T}\left( {{\bf{b}} + \varepsilon {\bf{x}}} \right)} \right\}} \right)
%\end{array} \right\}}.
%\end{equation}

By observing that
\begin{align}
&\sum\limits_{\bf{y}} {E\left( {\exp \left\{ { - {{\bf{y}}^T}\left( {{\bf{b}} + \varepsilon {\bf{x}}} \right)} \right\}} \right) + E\left( {\exp \left\{ {{{\bf{y}}^T}\left( {{\bf{b}} + \varepsilon {\bf{x}}} \right)} \right\}} \right)}  = \nonumber \\
&2\left( {\prod\limits_{i = 1}^n {\exp \{ {b_i} + \varepsilon {x_i}\}  + \exp  - \{ {b_i} + \varepsilon {x_i}\} } } \right) \nonumber 
\end{align}
and interchanging the order of summation and minimization, we simplify the upper bound,
%
%It is possible to interchange the order of summation and minimization
%Interchanging the order of summation and minimization and observing that $\sum\limits_{\bf{y}} {E\left( {\exp \left\{ { - {{\bf{y}}^T}\left( {{\bf{b}} + \varepsilon {\bf{x}}} \right)} \right\}} \right) + E\left( {\exp \left\{ {{{\bf{y}}^T}\left( {{\bf{b}} + \varepsilon {\bf{x}}} \right)} \right\}} \right)}  = 2\left( {\prod\limits_{i = 1}^n {\exp \{ {b_i} + \varepsilon {x_i}\}  + \exp  - \{ {b_i} + \varepsilon {x_i}\} } } \right)$ results in, 
\begin{equation}
{P_e} \le \frac{1}{{{{\left( {\frac{c}{2}} \right)}^n}}}\mathop {\min }\limits_{\bf{b}} E\left( {\prod\limits_{i = 1}^n {\cosh ({b_i} + \varepsilon {x_i})} } \right).
\end{equation}
Now, setting $\bf{b}= b \bf{e}$ (which would loosen the bound) and using the identity $\prod\limits_{i = 1}^n {\cosh (b + \varepsilon {x_i})} = 
{\left( {\frac{{\cosh (2b) + \cosh (2\varepsilon )}}{2}} \right)^{n/2}}{\left( {\sqrt {\frac{{\cosh (b + \varepsilon)}}{{\cosh (b - \varepsilon)}}} } \right)^{{{\bf{e}}^T}{\bf{x}}}},$
the upper bound simplifies further to
%
%Using the identity, $\prod\limits_{i = 1}^n {\cosh (b + \varepsilon {x_i})}  = {\left( {\frac{{\cosh (2b) + \cosh (2\varepsilon )}}{2}} \right)^{n/2}}{\left( {\sqrt {\frac{{\cosh (b + e)}}{{\cosh (b - e)}}} } \right)^{{{\bf{e}}^T}{\bf{x}}}}$ further simplifies the upper bound as,
\begin{equation}
{P_e} \le \frac{{\mathop {\min }\limits_b E\left( {{{\left( {\frac{{\cosh (2b) + \cosh (2\varepsilon )}}{2}} \right)}^{n/2}}{{\left( {\sqrt {\frac{{\cosh (b + \varepsilon )}}{{\cosh (b - \varepsilon )}}} } \right)}^{{{\bf{e}}^T}{\bf{x}}}}} \right)}}{{{{\left( {\cosh (\varepsilon )} \right)}^n}}}. \nonumber
\end{equation}
We can finally expand the expectation within the minimization to produce the upper bound:
\begin{equation}
{P_{e,UB}} = \frac{1}{{{{\left( {\cosh (\varepsilon )} \right)}^n}}}\mathop {\min }\limits_b \left( \begin{array}{l}
{\left( {\frac{{\cosh (2b) + \cosh (2\varepsilon )}}{2}} \right)^{n/2}}\\
 \times \frac{{\sum\limits_{\bf{x}} {\exp \left\{ {\theta {{\bf{x}}^T}A{\bf{x}} - \beta {{\bf{e}}^T}{\bf{x}}} \right\}} }}{{\sum\limits_{\bf{x}} {\exp \left\{ {\theta {{\bf{x}}^T}A{\bf{x}} - \gamma {{\bf{e}}^T}{\bf{x}}} \right\}} }}
\end{array} \right) \nonumber
\end{equation}
or, by recognizing that the summations (over $\mathbf{x}$) have the form of the standard MRF partition function, we finally obtain
the upper bound,
\begin{equation}
{P_{e,UB}} = \frac{1}{{Z(\theta ,\gamma ){{\left( {\cosh (\varepsilon )} \right)}^n}}}\mathop {\min }\limits_b \left( \begin{array}{l}
{\left( {\frac{{\cosh (2b) + \cosh (2\varepsilon )}}{2}} \right)^{n/2}}\\
 \times Z(\theta ,\beta )
\end{array} \right), \nonumber
\end{equation}
where
%Recognizing that the summations (over $\mathbf{x}$) have the form of the standard MRF partition function, we rewrite the upper bound as, 
%
%\begin{equation}
%{P_{e,UB}} = \frac{1}{{Z(\theta ,\gamma ){{\left( {\cosh (\varepsilon )} \right)}^n}}}\mathop {\min }\limits_b \left( \begin{array}{l}
%{\left( {\frac{{\cosh (2b) + \cosh (2\varepsilon )}}{2}} \right)^{n/2}}\\
% \times Z(\theta ,\beta )
%\end{array} \right),
%\end{equation}
%where
$\beta  = \gamma  + \frac{1}{2}\log \left( {\frac{{\cosh (b - \varepsilon )}}{{\cosh (b + \varepsilon )}}} \right)$, $Z(\theta ,\beta ) = \sum\limits_{\bf{x}} {\exp \left\{ {\theta {{\bf{x}}^T}A{\bf{x}} - \beta {{\bf{e}}^T}{\bf{x}}} \right\}}$ and $Z(\theta ,\gamma ) = \sum\limits_{\bf{x}} {\exp \left\{ {\theta {{\bf{x}}^T}A{\bf{x}} - \gamma {{\bf{e}}^T}{\bf{x}}} \right\}}$.
\vspace{-0.3 cm}
\bibliographystyle{IEEEtran}
 \bibliography{vinay}

%\begin{align}
%\beta  = \gamma  + \frac{1}{2}\log \left( {\frac{{\cosh (b - \varepsilon )}}{{\cosh (b + \varepsilon )}}} \right) \nonumber
%\end{align}
%\begin{align}
%Z(\theta ,\beta ) = \sum\limits_{\bf{x}} {\exp \left\{ {\theta {{\bf{x}}^T}A{\bf{x}} - \beta {{\bf{e}}^T}{\bf{x}}} \right\}} \nonumber
%\end{align}
%\begin{align}
%Z(\theta ,\gamma ) = \sum\limits_{\bf{x}} {\exp \left\{ {\theta {{\bf{x}}^T}A{\bf{x}} - \gamma {{\bf{e}}^T}{\bf{x}}} \right\}}
%\end{align}
%\begin{thebibliography}{25}
%\bibitem{book_twitter}Howard, Philip N., and Muzammil M. Hussain. Democracy's Fourth Wave?: Digital Media and the Arab Spring. Oxford University Press, 2013.}
%
%\bibitem{relay_1}Nishimori, Hidetoshi. ``Statistical physics of spin glasses
%and information processing: an introduction'', volume
%111. Clarendon Press, 2001.
%
%\end{thebibliography}

\end{document}